# COMPARISON OF WIDE AND COMPACT FOURTH ORDER FORMULATIONS OF THE NAVIER-STOKES EQUATIONS

**Ercan Erturk***
Gebze Institute of Technology, Energy Systems Engineering Department,
Gebze, Kocaeli 41400, Turkey

In this study the numerical performances of wide and compact fourth order formulation of the steady 2-D incompressible Navier-Stokes equations will be investigated and compared with each other. The benchmark driven cavity flow problem will be solved using both wide and compact fourth order formulations and the numerical performances of both formulations will be presented and also the advantages and disadvantages of both formulations will be discussed.

## 1. INTRODUCTION

In Computational Fluid Dynamics (CFD) field of study, most second order accurate finite difference approaches use three point discretizations. Depending on the flow problem, if a higher spatial accuracy is desired then fourth order accuracy can be chosen. In order to achieve fourth order spatial accuracy, standard five point discretization can be used. With five point discretization, wide fourth order formulation of the Navier-Stokes formulation has disadvantages near the boundaries such that when a wide fourth order formulation is used the points adjacent to the boundaries have to be treated specially.

Another way of achieving fourth order spatial accuracy is to use compact fourth order formulations. Compact fourth order formulations provide fourth order spatial accuracy in a $3\times3$ stencil. The main advantage of this type of formulation is that it could be easily used at the points adjacent to the boundary without a complexity.

In the literature it is possible to find many studies that uses "compact high order finite difference approximations". These studies can be categorized into two depending on the approache used for obtaining the compact high order finite difference approximations. In one category, the compact high order finite difference approximations are basically generalizations of the Padé scheme and the studies of Lele, Visbal and Gaitonde and Zhong are examples for this type of "compact high order" studies. For this category of the compact high order finite difference approximations Zingg have compared the numerical efficiency of the non-compact and compact finite difference methods.

In the second category, basically the Taylor series and the derivatives of the governing equations are used in obtaining the compact high order finite difference approximations. As examples of this type of "compact high order" studies, Spotz and Carey [18] and Li *et al.* [13], Zhang [23], Dennis and Hudson [5], MacKinnon and Johnson [14] and Gupta *et al.* [11] have demonstrated the efficiency of compact fourth order formulations of the streamfunction and vorticity equations for uniform grids, and also Ge and Zhang [9] and Spotz and Carey [19] have applied the fourth order compact formulation to nonuniform grids. This study is intended to contribute to this type of "compact high order" literature.

* E-mail: ercanerturk@gyte.edu.tr
URL: http://www.cavityflow.com

Recently Erturk and Gokcol [7] have presented a new fourth order compact formulation. The uniqueness of this formulation is that the presented compact fourth order formulation of the Navier-Stokes equations are in the same form with the Navier-Stokes equations with additional coefficients. In this formulation if these coefficients are chosen as zero the Navier-Stokes equations are obtained. Therefore if a numerical code is written for the solution of the introduced compact fourth order formulation, the obtained numerical solution is spatially fourth order ($\mathcal{O}(\Delta x^4)$) accurate to the Navier-Stokes equations. In this code if the coefficients are chosen as zero then the solution of the introduced compact fourth order formulation is spatially second order ($\mathcal{O}(\Delta x^2)$) accurate to the Navier-Stokes equations. Therefore with this formulation, using a single equation one can either obtain a second accurate or a fourth order accurate numerical solution of the Navier-Stokes equations just by setting up some coefficients. Moreover, using the introduced formulation one can easily change an existing second order accurate numerical code to provide fourth order accuracy and to do this the only thing have to be done is to add some coefficients into the existing second order accurate numerical code. When these coefficients are added into a second order code to obtain fourth accurate solutions, the code will run slower because of the extra CPU work of evaluating these inserted coefficients. Erturk [8] have shown that this extra CPU work is slightly dependent on the numerical method used. Erturk [8] also showed that for the driven cavity flow when ADI method is used in order to achieve numerical solutions with the same level of convergence (defined as convergence to the same maximum percent change in flow variables), a fourth order accurate numerical code requires 1.8 times more CPU time than a second order accurate numerical code.

Using the Taylor series expansion, one can easily prove that both compact fourth order formulations and five point wide fourth order formulations indeed provide fourth order spatial accurate numerical solutions. In the literature, to the best of authors' knowledge, there is no study that compares the solutions of compact fourth order formulation and the solutions of wide fourth order formulation in terms of accuracy and numerical performance.

The aim of this study is then to solve the steady 2-D incompressible Navier-Stokes equations with fourth order accuracy using both compact fourth order formulation and wide fourth order formulation. First in order to compare the spatial accuracy of both formulations we will consider an analytical test case. Then in order to compare both formulations in terms of numerical performances, as a test case we will consider the benchmark driven cavity flow for different Reynolds numbers with different grid mesh and we will compare the CPU time and number of iterations needed for convergence for both formulations. We will also document the maximum allowable time step that could be used in both formulations as a sign of stability characteristics of the formulations. Finally we will document the advantages and disadvantages of each formulation.

## 2. FOURTH ORDER COMPACT FORMULATION

We consider the steady 2-D incompressible Navier-Stokes equations in streamfunction ($\psi$) and vorticity ($\omega$) formulation. In non-dimensional form they are given as

$$\frac{\partial^2\psi}{\partial x^2}+\frac{\partial^2\psi}{\partial y^2}=-\omega \tag{1}$$

$$\frac{1}{Re}\frac{\partial^2\omega}{\partial x^2}+\frac{1}{Re}\frac{\partial^2\omega}{\partial y^2}=\frac{\partial\psi}{\partial y}\frac{\partial\omega}{\partial x}-\frac{\partial\psi}{\partial x}\frac{\partial\omega}{\partial y} \tag{2}$$

where $x$ and $y$ are the Cartesian coordinates and $Re$ is the Reynolds number.

If we discretize equations (1) and (2) using three point central differencing, the following finite difference equations provide second order ($\mathcal{O}(\Delta x^2,\Delta y^2)$) accurate solutions

$$\psi_{xx}+\psi_{yy}=-\omega \tag{3}$$

$$\frac{1}{Re}\omega_{xx}+\frac{1}{Re}\omega_{yy}=\psi_y\omega_x-\psi_x\omega_y \tag{4}$$

where subscripts denote second order ($\mathcal{O}(\Delta x^2,\Delta y^2)$) three point central finite difference derivative approximations.

As explained in Erturk and Gokcol [7], with using second order ($\mathcal{O}(\Delta x^2,\Delta y^2)$) three point central differencing, the solution of the following finite difference equations are fourth order ($\mathcal{O}(\Delta x^4,\Delta x^2\Delta y^2,\Delta y^4)$) accurate to the Navier-Stokes equations (1) and (2).

$$\psi_{xx}+\psi_{yy}=-\omega+A \tag{5}$$

$$\frac{1}{Re}(1+B)\omega_{xx}+\frac{1}{Re}(1+C)\omega_{yy}=(\psi_y+D)\omega_x-(\psi_x+E)\omega_y+F \tag{6}$$

where

$$A=-\frac{\Delta x^2}{12}\omega_{xx}-\frac{\Delta y^2}{12}\omega_{yy}-\left(\frac{\Delta x^2}{12}+\frac{\Delta y^2}{12}\right)\psi_{xxyy}$$

$$B=-Re\frac{\Delta x^2}{6}\psi_{xy}+Re^2\frac{\Delta x^2}{12}\psi_y\psi_y$$

$$C=Re\frac{\Delta y^2}{6}\psi_{xy}+Re^2\frac{\Delta y^2}{12}\psi_x\psi_x$$

$$D=\left(\frac{\Delta x^2}{12}+\frac{\Delta y^2}{12}\right)\psi_{xxy}-Re\frac{\Delta x^2}{12}\psi_y\psi_{xy}+Re\frac{\Delta y^2}{12}\psi_x\psi_{yy}$$

$$E=\left(\frac{\Delta x^2}{12}+\frac{\Delta y^2}{12}\right)\psi_{xyy}-Re\frac{\Delta x^2}{12}\psi_y\psi_{xx}+Re\frac{\Delta y^2}{12}\psi_x\psi_{xy}$$

$$F=\left(\frac{\Delta x^2}{12}+\frac{\Delta y^2}{12}\right)\psi_y\omega_{xyy}-\left(\frac{\Delta x^2}{12}+\frac{\Delta y^2}{12}\right)\psi_x\omega_{xxy}-\frac{\Delta x^2}{6}\psi_{xx}\omega_{xy}$$
$$+\frac{\Delta y^2}{6}\psi_{yy}\omega_{xy}+Re\left(\frac{\Delta x^2}{12}+\frac{\Delta y^2}{12}\right)\psi_x\psi_y\omega_{xy}-\left(\frac{\Delta x^2}{12}-\frac{\Delta y^2}{12}\right)\omega_x\omega_y$$

$$-\frac{1}{Re}\left(\frac{\Delta x^2}{12}+\frac{\Delta y^2}{12}\right)\omega_{xxyy} \qquad (7)$$

We note that equations (5) and (6) are in the same form with equations (3) and (4) except with additional coefficients $A$, $B$, $C$, $D$, $E$ and $F$. In equations (5) and (6) if the coefficients $A$, $B$, $C$, $D$, $E$ and $F$ are chosen as zero, we identically obtain equations (3) and (4). Therefore the solution of equations (5) and (6)become second order accurate ($\mathcal{O}(\Delta x^2,\Delta y^2)$) to the Navier-Stokes equations when the coefficients are chosen as zero. On the other hand, if the coefficients $A$, $B$, $C$, $D$, $E$ and $F$ in equations (5) and (6) are calculated as they are defined in equation (7), then the solution of these equations ((5) and (6)) are fourth order accurate ($\mathcal{O}(\Delta x^4,\Delta x^2\Delta y^2,\Delta y^4)$) to the Navier-Stokes equations. As demonstrated in Erturk [8], equations (5) and (6) provide easy way of obtaining either second order or fourth order spatial accurate solutions of the Navier-Stokes simply just by using the appropriate coefficients.

In the limit when $\Delta x\to 0$ and $\Delta y\to 0$, the finite difference derivative approximations in equations (5), (6) and (7) could be written as partial derivatives as the following

$$\frac{\partial^2\psi}{\partial x^2}+\frac{\partial^2\psi}{\partial y^2}=-\omega+A \qquad (8)$$

$$\frac{1}{Re}(1+B)\frac{\partial^2\omega}{\partial x^2}+\frac{1}{Re}(1+C)\frac{\partial^2\omega}{\partial y^2}=\left(\frac{\partial\psi}{\partial y}+D\right)\frac{\partial\omega}{\partial x}-\left(\frac{\partial\psi}{\partial x}+E\right)\frac{\partial\omega}{\partial y}+F \qquad (9)$$

where

$$A=-\frac{\Delta x^2}{12}\frac{\partial^2\omega}{\partial x^2}-\frac{\Delta y^2}{12}\frac{\partial^2\omega}{\partial y^2}-\left(\frac{\Delta x^2}{12}+\frac{\Delta y^2}{12}\right)\frac{\partial^4\psi}{\partial x^2\partial y^2}$$

$$B=-Re\frac{\Delta x^2}{6}\frac{\partial^2\psi}{\partial x\partial y}+Re^2\frac{\Delta x^2}{12}\frac{\partial\psi}{\partial y}\frac{\partial\psi}{\partial y}$$

$$C=Re\frac{\Delta y^2}{6}\frac{\partial^2\psi}{\partial x\partial y}+Re^2\frac{\Delta y^2}{12}\frac{\partial\psi}{\partial x}\frac{\partial\psi}{\partial x}$$

$$D=\left(\frac{\Delta x^2}{12}+\frac{\Delta y^2}{12}\right)\frac{\partial^3\psi}{\partial x^2\partial y}-Re\frac{\Delta x^2}{12}\frac{\partial\psi}{\partial y}\frac{\partial^2\psi}{\partial x\partial y}+Re\frac{\Delta y^2}{12}\frac{\partial\psi}{\partial x}\frac{\partial^2\psi}{\partial y^2}$$

$$E=\left(\frac{\Delta x^2}{12}+\frac{\Delta y^2}{12}\right)\frac{\partial^3\psi}{\partial x\partial y^2}-Re\frac{\Delta x^2}{12}\frac{\partial\psi}{\partial y}\frac{\partial^2\psi}{\partial x^2}+Re\frac{\Delta y^2}{12}\frac{\partial\psi}{\partial x}\frac{\partial^2\psi}{\partial x\partial y}$$

$$F=\left(\frac{\Delta x^2}{12}+\frac{\Delta y^2}{12}\right)\frac{\partial\psi}{\partial y}\frac{\partial^3\omega}{\partial x\partial y^2}-\left(\frac{\Delta x^2}{12}+\frac{\Delta y^2}{12}\right)\frac{\partial\psi}{\partial x}\frac{\partial^3\omega}{\partial x^2\partial y}-\frac{\Delta x^2}{6}\frac{\partial^2\psi}{\partial x^2}\frac{\partial^2\omega}{\partial x\partial y}$$
$$+\frac{\Delta y^2}{6}\frac{\partial^2\psi}{\partial y^2}\frac{\partial^2\omega}{\partial x\partial y}+Re\left(\frac{\Delta x^2}{12}+\frac{\Delta y^2}{12}\right)\frac{\partial\psi}{\partial x}\frac{\partial\psi}{\partial y}\frac{\partial^2\omega}{\partial x\partial y}-\left(\frac{\Delta x^2}{12}-\frac{\Delta y^2}{12}\right)\frac{\partial\omega}{\partial x}\frac{\partial\omega}{\partial y}$$
$$-\frac{1}{Re}\left(\frac{\Delta x^2}{12}+\frac{\Delta y^2}{12}\right)\frac{\partial^4\omega}{\partial x^2\partial y^2} \qquad (10)$$

As it is described briefly in Erturk and Gokcol [7], the numerical solutions of equations (8) and (9) are fourth order accurate ($\mathcal{O}(\Delta x^4, \Delta x^2 \Delta y^2, \Delta y^4)$) to the Navier-Stokes equations (1) and (2), ***strictly*** provided that second order central discretizations ($\mathcal{O}(\Delta x^2, \Delta y^2)$) are used and also ***strictly*** provided that a uniform grid mesh with $\Delta x$ and $\Delta y$ is used.

One thing is important to note here that, the Navier-Stokes equations (1) and (2) are a subset of equations (8) and (9) such that, when the coefficients $A$, $B$, $C$, $D$, $E$ and $F$ in equations (8) and (9) are chosen as zero, we obtain the Navier-Stokes equations (1) and (2). Since equations (8) and (9) and the Navier-Stokes equations (1) and (2) are in the same form, any numerical method suitable for the Navier-Stokes equations (1) and (2) can easily be applied to solve equations (8) and (9).

Moreover, if we have an existing code that solve the NS equations (1) and (2) with second order accuracy, we can easily alter the existing code by adding the coefficients $A$, $B$, $C$, $D$, $E$ and $F$ defined in equation (10) into the code, then the existing second order accurate code will turn into a fourth order accurate code. Therefore a single numerical code for the solution of equations (8) and (9) can provide both second order and fourth order spatial accuracy just by setting some coefficients.

## 3. NUMERICAL SOLUTION METHODOLOGY

We will numerically solve both the NS equations (1) and (2) and the introduced compact fourth order formulation of the Navier-Stokes equations (8) and (9). Both of these equation sets are nonlinear and therefore they need to be solved in an iterative manner. In order to have an iterative numerical algorithm we assign pseudo time derivatives to these equations, such that as an example for equations (8) and (9) we obtain the following

$$\frac{\partial \psi}{\partial t} = \frac{\partial^2 \psi}{\partial x^2} + \frac{\partial^2 \psi}{\partial y^2} + \omega - A \tag{11}$$

$$\frac{\partial \omega}{\partial t} = \frac{1}{Re}(1+B)\frac{\partial^2 \omega}{\partial x^2} + \frac{1}{Re}(1+C)\frac{\partial^2 \omega}{\partial y^2} - \left(\frac{\partial \psi}{\partial y} + D\right)\frac{\partial \omega}{\partial x} + \left(\frac{\partial \psi}{\partial x} + E\right)\frac{\partial \omega}{\partial y} - F \tag{12}$$

We solve these equations (11) and (12) in the pseudo time domain until the solutions converge to steady state.

For the solution method in the pseudo time, we decided to use the Alternating Direction Implicit (ADI) method. ADI method is a very widely used numerical method and in this method a two dimensional problem is solved in two sweeps while solving the equation implicitly in one dimension in each sweep. The reader is referred to [15], [1] and [4] for details.

### *3.1 Compact Fourth Order Solution Methodology*

In order to obtain solutions of the fourth order compact formulation, when we apply the ADI method to solve equations (8) and (9), first we solve the streamfunction equation and in order to do this we first solve the following tri-diagonal system in $x$-

direction

$$\left(1-\frac{\Delta t}{2}\delta_{xx}\right)\psi^{n+\frac{1}{2}}=\psi^{n}+\frac{\Delta t}{2}\delta_{yy}\psi^{n}+\frac{\Delta t}{2}\omega-\frac{\Delta t}{2}A \tag{13}$$

then we solve the following tri-diagonal system in $y$-direction

$$\left(1-\frac{\Delta t}{2}\delta_{yy}\right)\psi^{n+1}=\psi^{n+\frac{1}{2}}+\frac{\Delta t}{2}\delta_{xx}\psi^{n+\frac{1}{2}}+\frac{\Delta t}{2}\omega-\frac{\Delta t}{2}A \tag{14}$$

After this we solve the vorticity equation and in order to do this we first solve the following tri-diagonal system in $x$-direction

$$\left(1-\frac{\Delta t}{2}\frac{1}{Re}(1+B)\delta_{xx}+\frac{\Delta t}{2}(\psi_{y}+D)\delta_{x}\right)\omega^{n+\frac{1}{2}}=\omega^{n}$$
$$+\frac{\Delta t}{2}(1+C)\frac{1}{Re}\delta_{yy}\omega^{n}+\frac{\Delta t}{2}(\psi_{x}+E)\delta_{y}\omega^{n}-\frac{\Delta t}{2}F \tag{15}$$

then we solve the following tri-diagonal system in $y$-direction

$$\left(1-\frac{\Delta t}{2}\frac{1}{Re}(1+C)\delta_{yy}-\frac{\Delta t}{2}(\psi_{x}+E)\delta_{y}\right)\omega^{n+1}=\omega^{n+\frac{1}{2}}$$
$$+\frac{\Delta t}{2}\frac{1}{Re}(1+B)\delta_{xx}\omega^{n+\frac{1}{2}}-\frac{\Delta t}{2}(\psi_{y}+D)\delta_{xx}\omega^{n+\frac{1}{2}}-\frac{\Delta t}{2}F \tag{16}$$

In equations (13), (14), (15) and (16), the superscripts denote the iteration time level, $\delta_{xx}$ and $\delta_{yy}$ denote the second derivative three point central finite difference operators, and similarly $\delta_{x}$ and $\delta_{y}$ denote the first derivative three point central finite difference operators in $x$- and $y$-direction respectively, for example

$$\delta_{x}\theta=\frac{\theta_{i+1,j}-\theta_{i-1,j}}{2\Delta x}+\mathcal{O}(\Delta x^{2})$$
$$\delta_{xx}\theta=\frac{\theta_{i+1,j}-2\theta_{i,j}+\theta_{i-1,j}}{\Delta x^{2}}+\mathcal{O}(\Delta x^{2}) \tag{17}$$

where $i$ and $j$ are the grid index and $\theta$ denote any differentiable quantity.

We note that we use the Thomas algorithm for the solution of these tri-diagonal systems . Using the above numerical procedure defined in equations ((13), (14), (15) and (16)) we solve equations (8) and (9) and obtain spatially fourth order accurate solutions of the Navier-Stokes equations.

### *3.2 Wide Fourth Order Solution Methodology*

In order to obtain solutions of the fourth order wide formulation, when we apply the ADI method to solve the NS equations (1) and (2), similarly first we solve the streamfunction equation and in order to do this we first solve the following penta-diagonal system in $x$-direction

$$\left(1-\frac{\Delta t}{2}\delta_{xx}\right)\psi^{n+\frac{1}{2}}=\psi^{n}+\frac{\Delta t}{2}\delta_{yy}\psi^{n}+\frac{\Delta t}{2}\omega \tag{18}$$

then we solve the following penta-diagonal system in $y$-direction

$$\left(1-\frac{\Delta t}{2}\delta_{yy}\right)\psi^{n+1}=\psi^{n+\frac{1}{2}}+\frac{\Delta t}{2}\delta_{xx}\psi^{n+\frac{1}{2}}+\frac{\Delta t}{2}\omega \tag{19}$$

Then, similarly, we solve the vorticity equation and in order to do this we first solve the following penta-diagonal system in $x$-direction

$$\left(1-\frac{\Delta t}{2}\frac{1}{Re}\delta_{xx}+\frac{\Delta t}{2}\psi_y\delta_x\right)\omega^{n+\frac{1}{2}}=\omega^n+\frac{\Delta t}{2}\frac{1}{Re}\delta_{yy}\omega^n+\frac{\Delta t}{2}\psi_x\delta_y\omega^n \tag{20}$$

then we solve the following penta-diagonal system in $y$-direction

$$\left(1-\frac{\Delta t}{2}\frac{1}{Re}\delta_{yy}-\frac{\Delta t}{2}\psi_x\delta_y\right)\omega^{n+1}=\omega^{n+\frac{1}{2}}+\frac{\Delta t}{2}\frac{1}{Re}\delta_{xx}\omega^{n+\frac{1}{2}}-\frac{\Delta t}{2}\psi_y\delta_{xx}\omega^{n+\frac{1}{2}} \tag{21}$$

We note that, in equations (18), (19), (20) and (21) the second and first derivative operators $\delta_{xx}$, $\delta_{yy}$, $\delta_x$ and $\delta_y$ denote five point central finite difference operators in $x$- and $y$-direction respectively, for example

$$\delta_x\theta=\frac{-\theta_{i+2,j}+8\theta_{i+1,j}-8\theta_{i-1,j}+\theta_{i-2,j}}{12\Delta x}+\mathcal{O}(\Delta x^4)$$

$$\delta_{xx}\theta=\frac{-\theta_{i+2,j}+16\theta_{i+1,j}-30\theta_{i,j}+16\theta_{i-1,j}-\theta_{i-2,j}}{12\Delta x^2}+\mathcal{O}(\Delta x^4) \tag{22}$$

where $i$ and $j$ are the grid index and $\theta$ denote any differentiable quantity.

We note that, in numerical solution of the wide formulation we use the modified Thomas algorithm for the solution of the penta-diagonal systems. Following the above numerical procedure defined in equations (18), (19), (20) and (21) we obtain spatially fourth order accurate solutions of the Navier-Stokes equations.

### *3.3 Boundary Conditions*

The compact formulations require a $3\times 3$ stencil and therefore this formulation can be easily used at the first set of grid points near the boundaries. However, wide formulations can not be applied directly to the first set of grid points near the boundaries. We are going to compare the numerical solutions of wide and compact formulations, therefore in order to isolate the effect of the boundary conditions on both formulations we would like to use the same boundary conditions in both cases, so that the effect of the boundary conditions will be the same in both cases.

Störtkuhl *et al.* [21] have presented an analytical asymptotic solution near the corners of the cavity and using finite element bilinear shape functions they also have presented a singularity removed boundary condition for vorticity at the corner grid points as well as at the wall grid points. In this study, we follow Störtkuhl *et al.* [21] and calculate the vorticity values at the wall grid points (circle points in Figure 1-a) as the following

$$\omega_{\mathrm{b}}=-\frac{9V}{2\Delta h}-\frac{3}{2\Delta h^2}(\psi_{\mathrm{d}}+\psi_{\mathrm{e}}+\psi_{\mathrm{f}})-\frac{1}{8}(2\omega_{\mathrm{a}}+2\omega_{\mathrm{c}}+\omega_{\mathrm{d}}+4\omega_{\mathrm{e}}+\omega_{\mathrm{f}}) \tag{23}$$

where $V$ is the speed of the wall which is equal to 1 for the moving top wall and equal to 0 for the three stationary walls. The grid points $a$, $b$, $c$, $d$, $e$ and $f$ are shown in Figure 1-b.

For corner points, we again follow Störtkuhl *et al.* [21] and calculate the vorticity values at the corner grid points (diamond point in Figure 1-a) as the following

$$\omega_{\mathrm{b}} = -\frac{9V}{2\Delta h} - \frac{3}{\Delta h^2}\psi_{\mathrm{f}} - \frac{1}{4}(2\omega_{\mathrm{c}} + \omega_{\mathrm{f}} + 2\omega_{\mathrm{e}}) \tag{24}$$

where again $V$ is equal to 1 for the upper two corners and it is equal to 0 for the bottom two corners. The grid points $b$, $c$, $e$ and $f$ are also shown in Figure 1-c. The reader is referred to Störtkuhl *et al.* [21] for details on the boundary conditions

In this study, for the first set of grid points adjacent to the wall we decided to use the Computational Boundary Method (CBM). For details on the CBM, the reader is referred to Huang [12], Yang [22], Gresho [10] and Spotz [20]. Using a fourth order one sided approximation for the velocity on the wall we obtain the following expression

$$\pm V_w = \left.\frac{\partial \psi}{\partial n}\right|_0 = \frac{-25\psi_0 + 48\psi_1 - 36\psi_2 + 16\psi_3 - 3\psi_4}{12h} + \mathcal{O}(h^4) \tag{25}$$

where $V_w$ denotes the wall velocity, $n$ is the normal wall direction, $h$ is the grid spacing, 0 denotes the grid points on the wall, 1 denotes the first set of grid points adjacent to the wall and similarly 2, 3 and 4 denotes the second, third and fourth set of grid points adjacent to the wall respectively. From this relation we can calculate the streamfunction at the grid points near the boundary (rectangle points in Figure 1-a) as the following

$$\psi_1 = \frac{\pm 12hV_w + 25\psi_0 + 36\psi_2 - 16\psi_3 + 3\psi_4}{48} \tag{26}$$

In order to calculate the vorticity values at these grid points, we used the streamfunction equation (1). Using a five point wide fourth order formulation for the streamfunction equation, the vorticity at the first set of grid points adjacent to the boundary (rectangle points in Figure 1-a) is calculated as the following

$$\begin{aligned}\omega_{i,j} = {} & \frac{-\psi_{i+2,j} + 16\psi_{i+1,j} - 30\psi_{i,j} + 16\psi_{i-1,j} - \psi_{i-2,j}}{12\Delta x^2} \\ & + \frac{-\psi_{i,j+2} + 16\psi_{i,j+1} - 30\psi_{i,j} + 16\psi_{i,j-1} - \psi_{i,j-2}}{12\Delta y^2} + \mathcal{O}(\Delta x^4, \Delta y^4)\end{aligned} \tag{27}$$

We note that, this approximation require the streamfunction values at grid points outside the computational domain (hexagon points in Figure 1-a). In order to find the streamfunction values at these grid points we use the following fourth order relation

$$\pm V_w = \left.\frac{\partial \psi}{\partial n}\right|_0 = \frac{-3\psi_{-1} - 10\psi_0 + 18\psi_1 - 6\psi_2 + \psi_3}{12h} + \mathcal{O}(h^4) \tag{28}$$

where -1 denotes the grid point outside the computational domain. From this relation we calculate the value of streamfunction at the exterior grid points ($\psi_{-1}$) and use it in equation (27) to calculate the vorticity at the grid points adjacent to the boundary.

At the first diagonal grid points near the corners (triangle point in Figure 1-a), we calculate the streamfunction and vorticity values, using Gauss-Seidel method on the five point fourth order wide formulation of the streamfunction equation (1) and vorticity equations (2) respectively.

We note that in both wide and compact formulations, we used the above described boundary conditions for the grid points on the wall and for the first set of grid points adjacent to the wall. Then we used the wide and compact formulations to obtain the numerical solutions at interior grid points shown in Figure 1-a.

## 4. RESULTS AND DISCUSSIONS

Following the numerical procedure described in the previous section we obtain spatially fourth order solutions of the Navier-Stokes equations obtained using both compact and wide formulations.

For the choice of time steps in solving the governing equations, we decided to use different time steps, $\Delta t$, for streamfunction and vorticity equations, as it was also done in Erturk [8]. If we solve the streamfunction and vorticity equations using three point second order accurate discretization using the ADI method, tri-diagonal matrices appear on the implicit LHS of the equations. In streamfunction equation the diagonal elements on the LHS matrices become $1+\frac{2\Delta t}{\Delta h^2}$, and in vorticity equations the diagonal elements on the LHS matrices become $1+\frac{1}{Re}\frac{2\Delta t}{\Delta h^2}$. We decided to choose different time steps for streamfunction and vorticity equations such that these different time steps would make the diagonal elements the same. Therefore we use $\Delta t = \alpha \Delta h^2$ for the streamfunction equation and similarly we use $\Delta t = \alpha Re \Delta h^2$ for the vorticity equation, where $\alpha$ is a coefficient we can choose. We solve both the wide and compact formulations using the same time steps.

In this study we would also like to document the maximum allowable time step $\Delta t$ that we can use in both formulation for convergence as a sign of numerical stability characteristics of both formulations. In order to this, we first solve the wide and compact formulations using the same time steps. Then we increase $\alpha$, i.e. increase the time step $\Delta t$, and solve the wide and compact formulations again. We continue doing this until at some $\Delta t$ the solution does not converge. Therefore we would document the maximum allowable $\Delta t$ for convergence for wide and compact formulation for a given Reynolds number and grid mesh.

In this study, in all of the cases considered, we start the iterations from a homogenous initial guess and continue until a certain condition of convergence is satisfied. As the measure of the convergence to the same level, one option is to use the residual of the equations as it was also used by Erturk, Corke and Gokcol [6]. However, we are actually solving two different equations, the NS equations (1) and (2) and also the fourth order equations (8) and (9), and we are trying to compare the numerical performance of these two different equations. Therefore the residual of these two different equations may not indicate the convergence to the same level. Alternatively, we can use the difference of the streamfunction and vorticity variables between two time steps as the measure of convergence for the two equations. However, the solutions of the two different equations are slightly different. Since the solutions are different, the difference of the streamfunction and vorticity variables between two time steps may not show the same convergence level for these equations also. Therefore in this study, as convergence criteria we decided to use the difference of the streamfunction and vorticity variables between two time steps normalized by the previous value of the corresponding variable, such that

$$\text{Residual}_\psi = \max\left(\left|\frac{\psi_{i,j}^{n+1}-\psi_{i,j}^{n}}{\psi_{i,j}^{n}}\right|\right)$$

$$\text{Residual}_\omega = \max\left(\left|\frac{\omega_{i,j}^{n+1}-\omega_{i,j}^{n}}{\omega_{i,j}^{n}}\right|\right) \tag{29}$$

These residuals provide an indication of the maximum percent change in $\psi$ and $\omega$ variables in each iteration step and in this study we let the iterations converge until both $\text{Residual}_\psi$ and $\text{Residual}_\omega$ are less than $10^{-8}$.

Using the Taylor series expansion, mathematically it is straight forward to show that the first leading truncation error term in both fourth order compact formulations and five point fourth order wide formulations is indeed fourth order. In order to document this numerically, we decided to compare the numerical results with a known analytical solution using a model problem introduced by Richards and Crane [16]. Inside the domain $(x, y) = ([0,1],[0,1])$ the following analytical solutions satisfy the Navier-Stokes equations (1) and (2).

$$\psi = \frac{y-x}{Re} - e^{(x+y)}$$

$$\omega = 2e^{(x+y)} \tag{30}$$

For this model problem, as the boundary conditions we decided to use the analytical solutions defined in equation (30) at the grid points on the boundaries and at the first set of grid points adjacent to the boundaries. This way we would be able to avoid any effect of numerical boundary condition approximations on the numerical solution and concentrate on the accuracy of the solution of both formulations in the interior domain for a given analytical values at the boundaries.

We note that, in equation (30) the vorticity is independent of $Re$ and the solution of the streamfunction at different $Re$ numbers looks almost the same in a contour plot, therefore for this model problem we only considered the case where $Re = 1000$. We solve this model problem with wide and compact fourth order formulations using different grid mesh with $\Delta h = 1/16, 1/32, 1/64, 1/128$ where $\Delta x = \Delta y = \Delta h$. In these fourth order solutions the average absolute difference between the exact solution ($\psi_{ex}, \omega_{ex}$) given in equation (30) and the numerical solution ($\psi_{i,j}, \omega_{i,j}$), defined as

$$E_\psi = \frac{\sum_{i,j}\left|\psi_{ex}(x_i, y_j) - \psi_{i,j}\right|}{N}$$

$$E_\omega = \frac{\sum_{i,j}\left|\omega_{ex}(x_i, y_j) - \omega_{i,j}\right|}{N} \tag{31}$$

where $N$ being the total number of grid points, should be proportional to $h^m$ where the power should be equal to $m = 4$. Since these average absolute differences should have the following form

$$E = C\Delta h^m \tag{32}$$

logarithm of both sides give

$$\log E = \log C + m \log \Delta h \tag{33}$$

In Table 1 we have tabulated the average absolute differences ($E_\psi$ and $E_\omega$) for wide and compact fourth order solutions for different grid mesh and also in Figure 2, these average absolute differences are plotted with respect to the grid spacing in a log-log scale. In Figure 2 the average absolute differences of the wide and compact formulations are shown by square and circle symbols respectively. In order to have an idea on how these average absolute differences change with respect to the grid spacing, two linear lines with blue and red colors with slopes of 4 are drawn in the same figure. From Figure 2 we can clearly see that both wide and compact fourth order formulations (shown by square and circle in the figure), indeed provide fourth order accurate solutions, such that for both formulations the slope between $\log E$ and $\log \Delta h$ is very close to $m \approx 4$.

Using the average absolute differences ($E_\psi$ and $E_\omega$) tabulated in Table 1, since the grid size is decreased by a factor of 2, we can calculate the convergence rate $m$ using the following formula

$$\frac{E\left(\Delta h = \Delta h_1\right)}{E\left(\Delta h = \dfrac{\Delta h_1}{2}\right)} = 2^m \tag{34}$$

The convergence rate $m$ for both wide and fourth order formulations is also tabulated in Table 1. From this table, we can again see that when the grid spacing is decreased progressively by half , the convergence rates of both formulations is very close to $m \approx 4$.

Next, we considered the benchmark driven cavity flow problem. For the driven cavity flow we obtain fourth order numerical solutions using both wide and compact fourth order formulations using the boundary conditions described in the previous section and also using both $128 \times 128$ and $256 \times 256$ grid mesh.

Figures 4, 5 and 6 show the streamfunction contours of the compact and wide formulation solutions for a variety of Reynolds numbers ($Re = 100$, $1000$, $2500$) obtained with using $256 \times 256$ grid points.

From these figures we can see that both compact and wide fourth order formulation solutions are very close to each other. The difference between them is most visible in the region of the center of the main circulation. In the literature, for the benchmark driven cavity flow, the value of the streamfunction and the vorticity at the center of the main circulation and also the location of this center is widely accepted as the flow parameters to compare the accuracy of the numerical solutions. In Table 2 we tabulated the value of the streamfunction and the vorticity at the center of the main circulation and the location of the center for the Reynolds numbers considered. In Table 2 we also have tabulated solutions found in the literature that we believe to be very accurate. Erturk and Gokcol [7] have presented compact fourth order ($\Delta h^4$) solutions obtained using a very fine grid mesh ($600 \times 600$). Botella and Peyret [3] have used a Chebyshev collocation method and obtained highly accurate spectral solutions for the driven cavity flow. Barragy and Carey [2] have used a *p*-type finite element scheme and presented highly accurate ($\Delta h^8$ order) solutions. Schreiber and Keller [17], have presented high-order ($\Delta h^8$ order in theory) solutions obtained using repeated Richardson extrapolation using the solutions obtained on different grid mesh sizes. Erturk, Corke and Gokcol [6] have also used repeated Richardson extrapolation and presented high-order ($\Delta h^6$ order in theory) solutions. From Table 2 we can see that the compact and wide formulation solutions are very close to each other and also all of the solutions agree well with the highly accurate solutions found in the literature.

In order to test the numerical performances of the wide and compact formulations, we then solve the benchmark driven cavity flow problem several times with different time steps $\Delta t$ by changing $\alpha$. In order to see the effect of the number of grids on the numerical performances of both formulations, we have tabulated the numerical performances of wide and compact formulations to achieve same level of convergence. Table 3 shows the CPU time and the number of iterations necessary for convergence of the wide and compact fourth order formulations for a $128\times 128$ grid mesh and also Table 4 shows the same for a $256\times 256$ grid mesh.

From Tables 3 and 4 we can see that, for the same Reynolds number and for the same time step ($\alpha$) both wide fourth order formulation and compact fourth order formulation converge to the same level in about the same number of iteration such that the ratio of the number of iteration necessary for convergence for wide formulation to the number of iteration necessary for convergence for compact formulation is approximately $\approx 1$. For wide and compact fourth order formulations, the convergence rate in the pseudo time is approximately the same.

From Tables 3 and 4 we also can see that, for the same Reynolds number and for the same time step ($\alpha$), the CPU time necessary for convergence of compact fourth order formulation is higher than the CPU time necessary for convergence of wide fourth order formulation. For the solution of the compact formulation, in each time step the coefficients $A$, $B$, $C$, $D$, $E$ and $F$ in equations (5) and (6) have to be calculated as they are defined in equation (7) and this would require a CPU time. The compact formulation requires the solution of tri-diagonal systems, and these tri-diagonal systems are solved efficiently using the Thomas algorithm. On the other hand, the wide formulation requires the solution of penta-diagonal systems. For these penta-diagonal systems we used the modified Thomas algorithm. The modified Thomas algorithm for penta-diagonal systems runs slower than the Thomas algorithm for tri-diagonal systems since it requires more mathematical instructions. Even though the modified Thomas algorithm runs slower, from Tables 3 and 4 we see that, for convergence the compact formulation requires more CPU time than that of the wide formulation. This shows that the extra CPU time necessary to calculate the coefficients $A$, $B$, $C$, $D$, $E$ and $F$ is higher than the extra CPU time necessary for the modified Thomas algorithm. For the same Reynolds number and for the same time step ($\alpha$), in terms of the CPU time necessary for convergence, the wide fourth order formulation is more advantageous than the compact forth order formulation such that the compact fourth order code requires almost $\approx 1.5$ more CPU time than the wide fourth order code.

For a numerical formulation, the largest time step that the numerical code would not diverge is an indicative of the numerical stability of the numerical formulation. In order to find the largest time step for the formulations, we progressively increased the time step ($\alpha$) with 0.01 and run the code several times for a given Reynolds number until the solution is no longer convergent such that we could not obtain a solution, i.e. the solution is divergent. In Tables 3 and 4 we can see that, in all cases, we can obtain numerical solution using larger time steps in compact formulation compared to that of wide formulation. This would indicate that the compact formulation is numerically more stable than the wide formulation and allows us to use larger time steps.

When running a code, one would like to use the maximum possible time step ($\Delta t$) for a faster convergence. In Table 3 and 4, comparing the CPU time of the compact formulation when maximum time step is considered ($\alpha = 1.3$) with the CPU time of the wide formulation when maximum time step is considered ($\alpha = 1.1$), we can see that when maximum possible time steps are used, the compact formulation runs approximately $\approx 1.3$ times slower than the wide formulation in terms of the CPU time.

## 2. CONCLUSIONS

In this study we have numerically solved the Navier-Stokes equations using both fourth order compact formulation and fourth order wide formulation and compared the numerical performances of both fourth order formulations.

Solving a model problem which has an exact analytical solution to N-S equations with several different grid mesh showed that the solution of both wide and compact formulations, indeed converge with fourth degree with respect to the grid spacing. Also solving the benchmark driven cavity flow showed that the numerical solution of both formulations are very close to each other and produce spatially very accurate results.

We see that both wide and compact formulations converge to a specified convergence level in almost the same number of iterations for a given Reynolds number and time step. In terms of number of iterations necessary for convergence, both formulations have the similar convergence rate.

For a given Reynolds number and time step, wide formulation requires less CPU time than the compact formulation. In terms of computing time, wide formulation runs faster and is advantageous over the compact formulation.

In compact formulation we were able to obtain convergence using larger time steps than the time steps we use in wide formulation. This would show that the compact formulation is numerically more stable than the wide formulation, i.e. allows us to use larger time steps. In terms of numerical stability compact formulation is advantageous over wide formulation.

We would like to point out that in order to use the wide formulation, the points adjacent to boundaries have to be treated specially and doing this adds a great complexity in coding stage. On the other hand, even though compact formulation does not have this complexity and can be easily used for the points adjacent to boundaries, in order to use a compact formulation one has to deal with the extra coefficients in the equations and this can be counted as the complexity of the compact formulation in coding stage. Therefore we can fairly say that both wide and compact formulation requires almost the same level of coding effort.

## REFERENCES

[1] Anderson DA, Tannehill JC, Pletcher RH. *Computational Fluid Mechanics and Heat Transfer*. McGraw-Hill, New York, 1984.

[2] Barragy E, Carey GF. Stream function-vorticity driven cavity solutions using *p* finite elements. *Computers and Fluids* 1997; **26**:453– 468.

[3] Botella O, Peyret R. Benchmark spectral results on the lid-driven cavity flow. *Computers and Fluids* 1998; **27**:421– 433.

[4] Chung TJ. *Computational Fluid Dynamics*. Cambridge University Press, Cambridge, 2002.

[5] Dennis SC, Hudson JD. Compact $h^4$ Finite Difference Approximations to Operators of Navier-Stokes Type, *Journal of Computational Physics* 1989; **85**:390–416.

[6] Erturk E, Corke TC, Gokcol C. Numerical Solutions of 2-D Steady Incompressible Driven Cavity Flow at High Reynolds Numbers, *International Journal for Numerical Methods in Fluids* 2005; **48**:747–774.

[7] Erturk E, Gokcol C. Fourth Order Compact Formulation of Navier-Stokes Equations and Driven Cavity Flow at High Reynolds Numbers, *International Journal for Numerical Methods in Fluids* 2006; **50**:421–436.

[8] Erturk E, Numerical Performance of Compact Fourth-Order Formulation of the Navier–Stokes Equations, *Communications in Numerical Methods in Engineering*, **In Press.**

[9] Ge L, Zhang J. High Accuracy Iterative Solution of Convection Diffusion Equation with Boundary Layers on Nonuniform Grids, *Journal of Computational Physics* 2001; **171**:560–578.

[10] Gresho PM. Incompressible Fluid Dynamics: Some Fundamental Formulation Issues, *Annual Review of Fluid Mechanics* 1991; **23**:413–453

[11] Gupta MM, Manohar RP, Stephenson JW. A Single Cell High Order Scheme for the Convection-Diffusion Equation with Variable Coefficients, *International Journal for Numerical Methods in Fluids* 1984; **4**:641–651.

[12] Huang H, Yang H. The Computational Boundary Method for Solving the Navier-Stokes Equations, Institute of Applied Mathematics and Statistics Technical Report 1990. No: Iam90-05, http://www.iam.ubc.ca/pub/techreports/.

[13] Li M, Tang T, Fornberg B. A Compact Forth-Order Finite Difference Scheme for the Steady Incompressible Navier-Stokes Equations, *International Journal for Numerical Methods in Fluids* 1995; **20**:1137–1151.

[14] MacKinnon RJ, Johnson RW. Differential-Equation-Based Representation of Truncation Errors for Accurate Numerical Simulation, *International Journal for Numerical Methods in Fluids* 1991; **13**:739–757.

[15] Peaceman DW, Rachford Jr. HH. The numerical solution of parabolic and elliptic differential equations, *Journal of the Society for Industrial and Applied Mathematics* 1955; **3**:28–41.

[16] Richards CW, Crane CM. The accuracy of finite difference schemes for the numerical solution of the Navier-Stokes equations, *Applied Mathematical Modelling* 1979; **3**:205–211.

[17] Schreiber R, Keller HB. Driven cavity flows by efficient numerical techniques. *Journal of Computational Physics* 1983; **49**:310 –333.

[18] Spotz WF, Carey GF. High-Order Compact Scheme for the Steady Streamfunction Vorticity Equations, *International Journal for Numerical Methods in Engineering* 1995; **38**:3497–3512.

[19] Spotz WF, Carey GF. Formulation and Experiments with High-Order Compact

Schemes for Nonuniform Grids, *International Journal of Numerical Methods for Heat and Fluid Flow* 1998; **8**:288–303.
[20] Spotz WF. Accuracy and Performance of Numerical Wall Boundary Conditions For Steady, 2D, Incompressible Streamfunction Vorticity, *International Journal for Numerical Methods in Fluids* 1998; **28**:737–757.
[21] Störtkuhl T, Zenger C, Zimmer S. An Asymptotic Solution for the Singularity at the Angular Point of the Lid Driven Cavity, *International Journal of Numerical Methods for Heat and Fluid Flow* 1994; **4**:47–59.
[22] Yang HH. The Computational Boundary Method for Solving the Incompressible Flows, *Applied Mathematics Letters* 1993; **6**:3–7.
[23] Zhang J. An Explicit Fourth-Order Compact Finite Difference Scheme for Three-Dimensional Convection-Diffusion Equation, *Communications in Numerical Methods in Engineering* 1998; **14**:209–218.

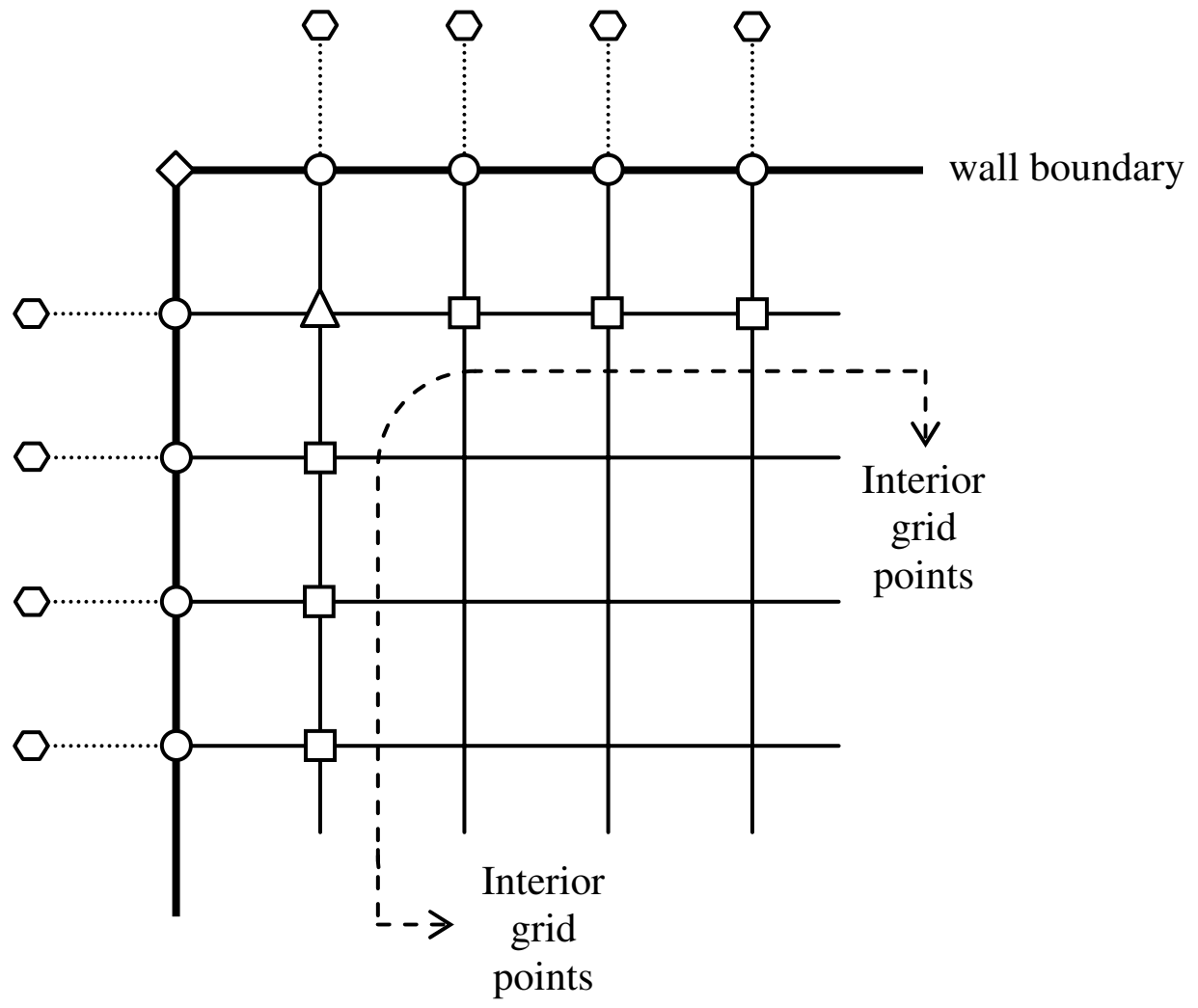

a) Grid points adjacent to wall

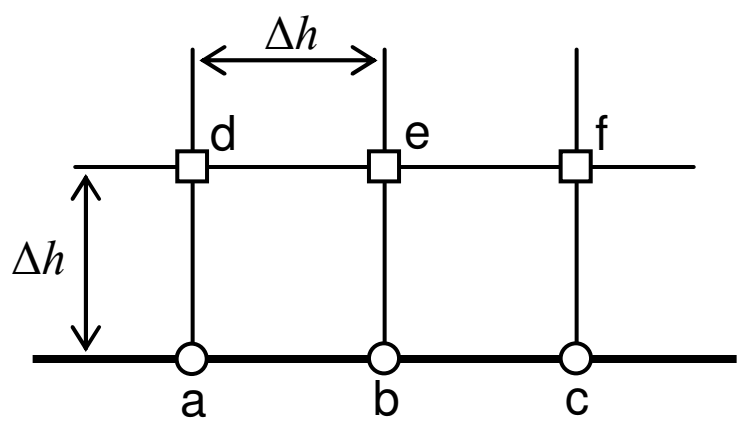

b) Grid points at the wall

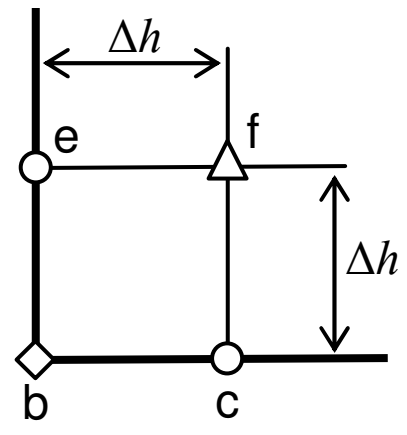

c) Grid points at the corners

Figure 1)   Grid points at boundaries

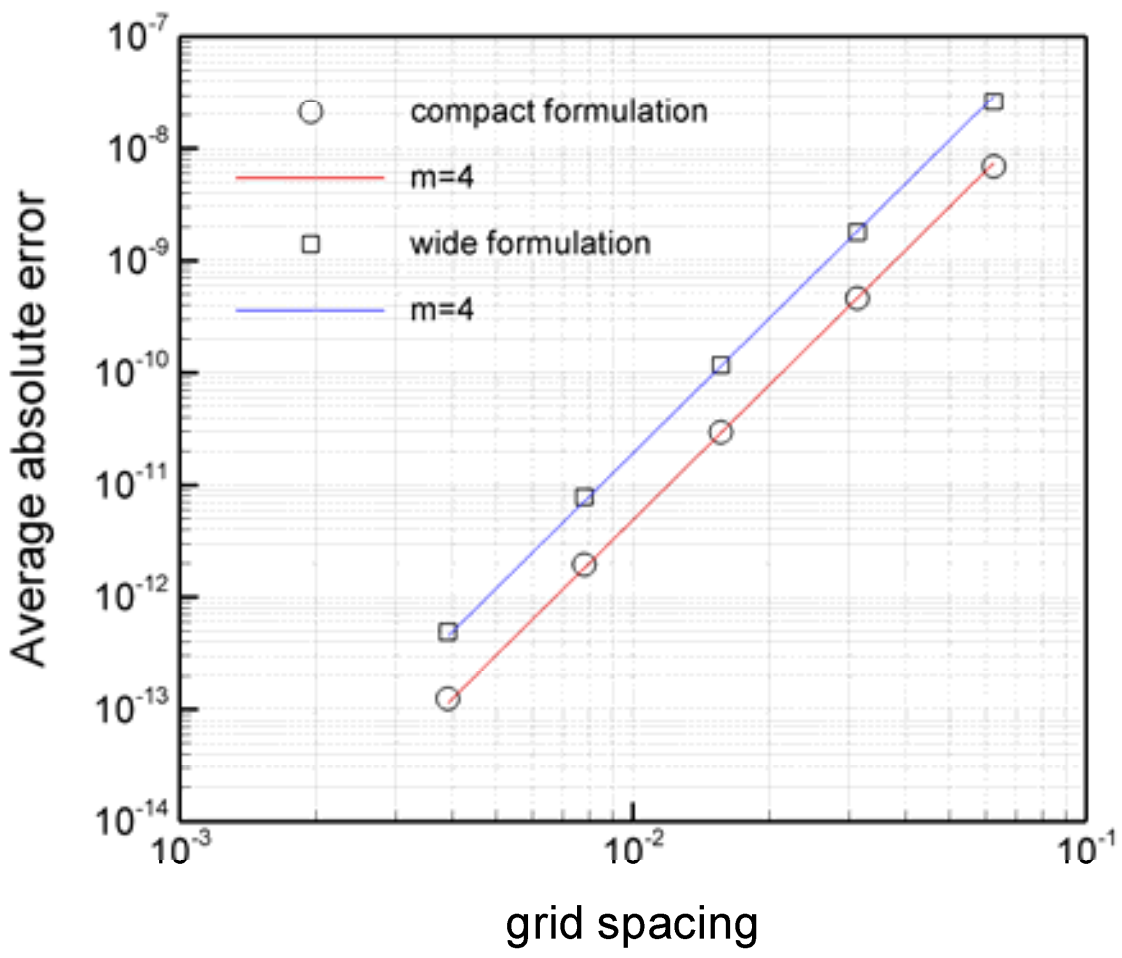

a ) streamfunction ( $E_{\psi}$ )

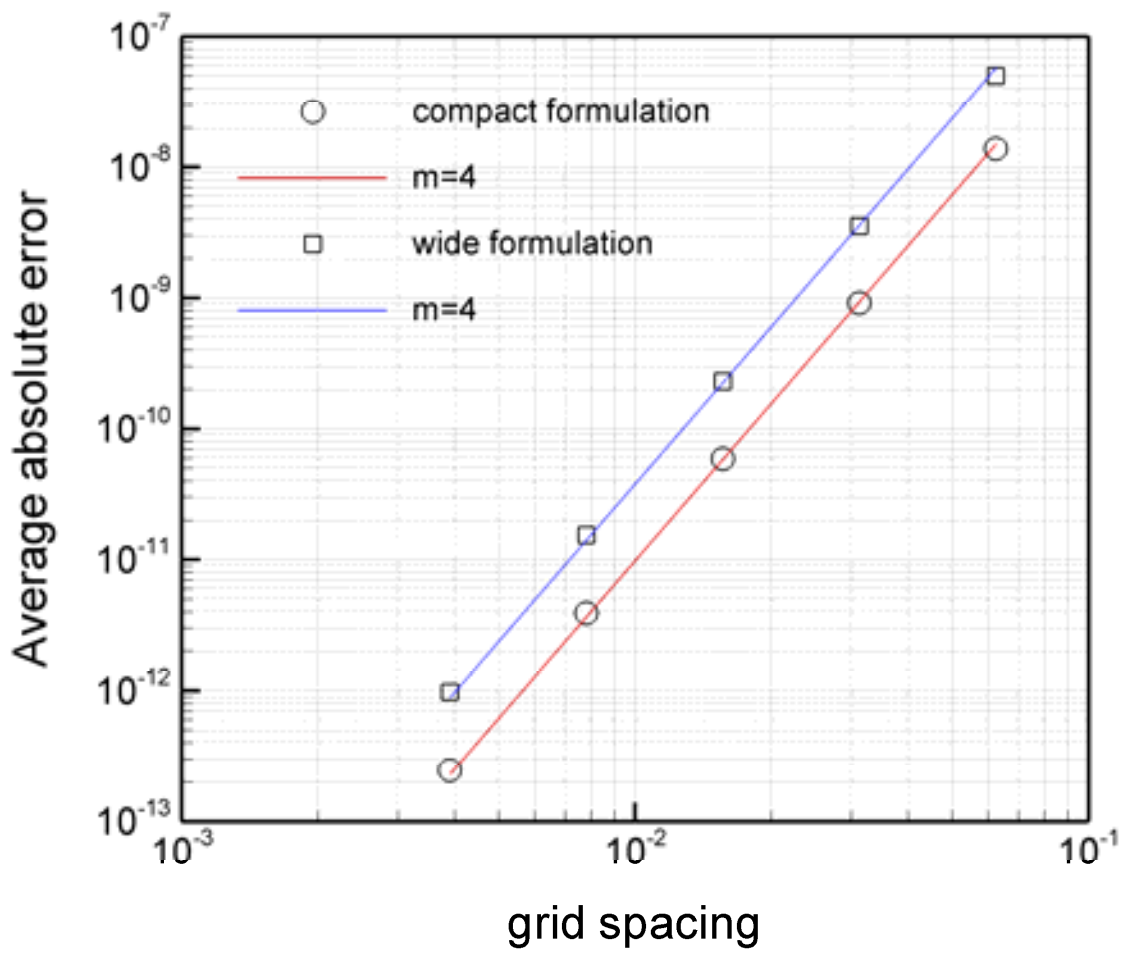

b) vorticity ( $E_{\omega}$ )

Figure 2) Average absolute errors in streamfunction and vorticity variables in fourth order wide and compact formulations with respect to grid spacing

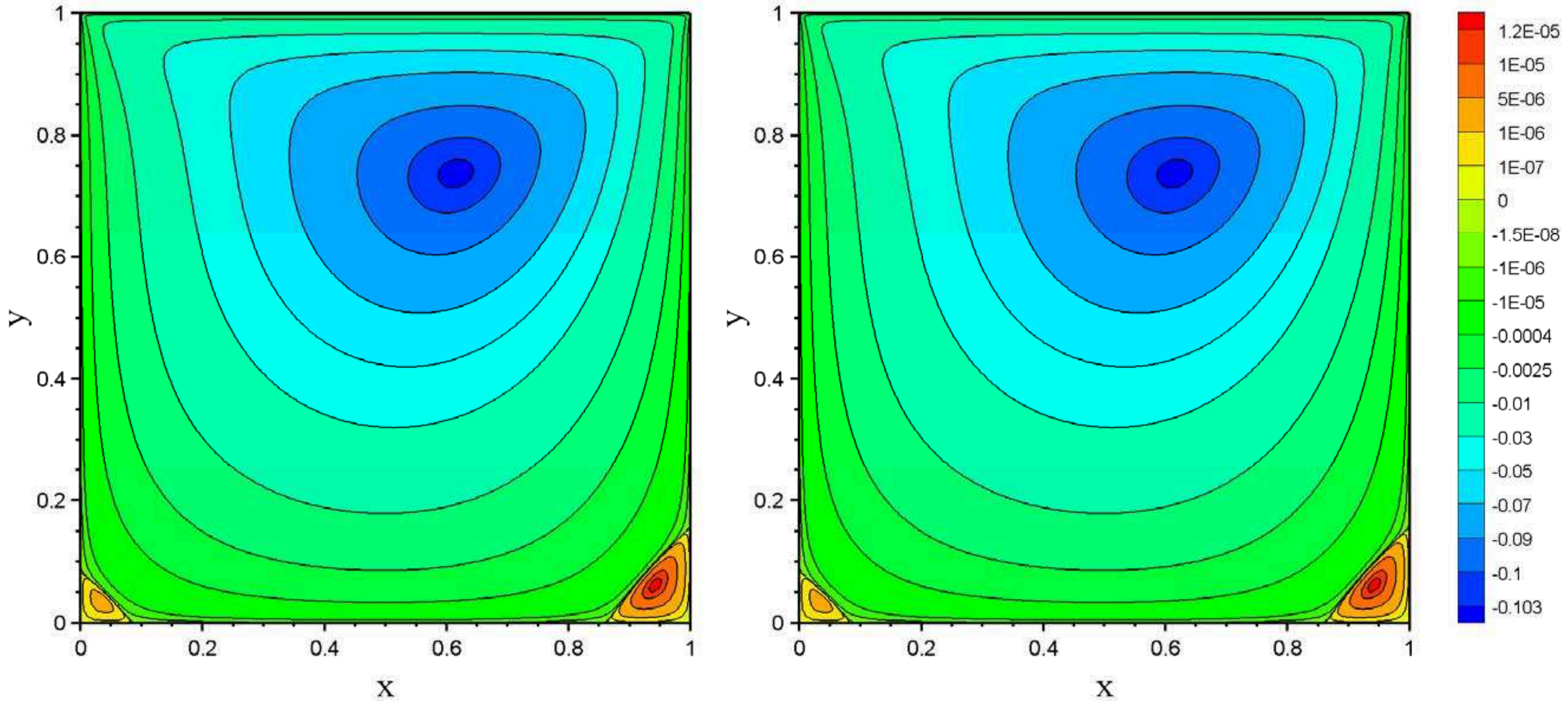

a) Compact solution

b) Wide solution

Figure 3) Fourth order streamfunction contours of driven cavity flow, $Re = 100$

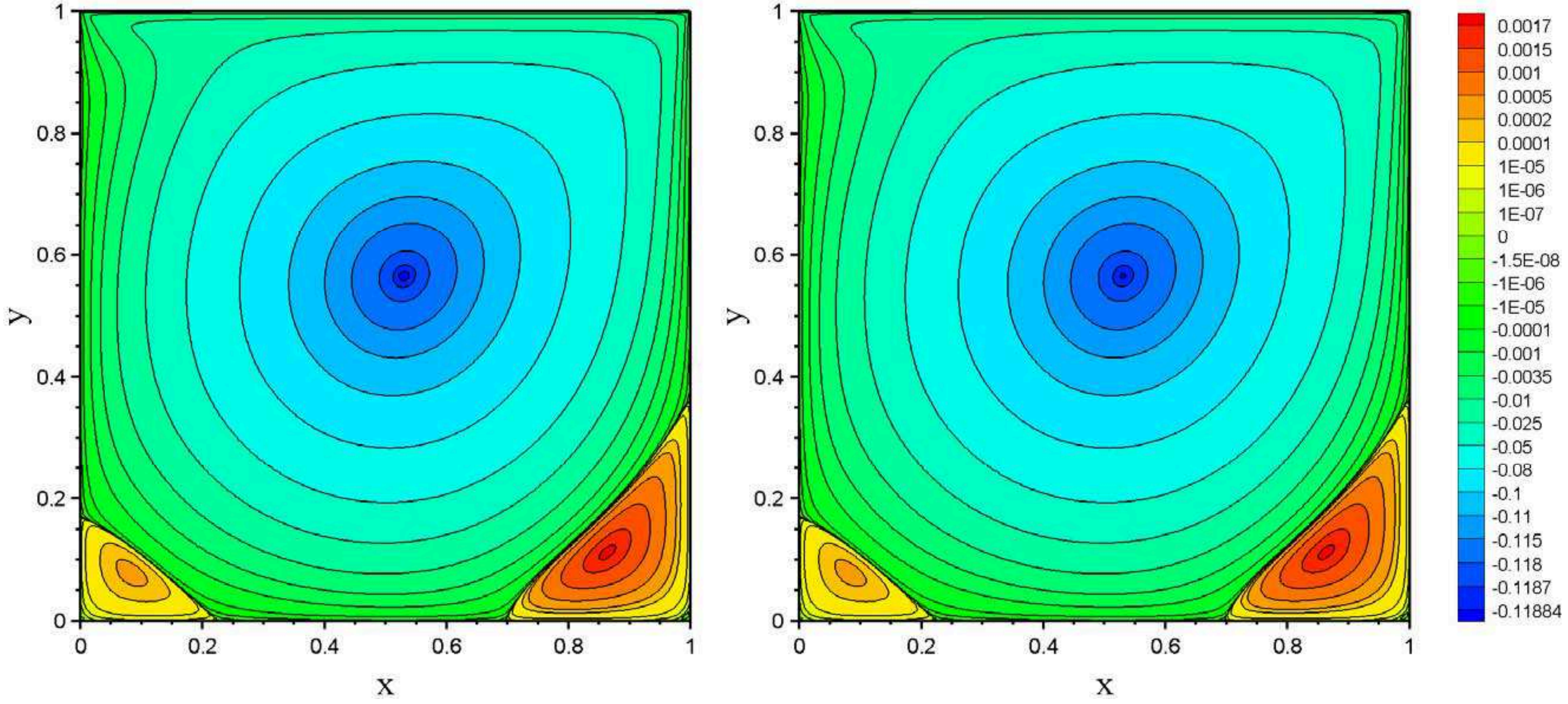

a) Compact solution

b) Wide solution

Figure 4) Fourth order streamfunction contours of driven cavity flow, $Re = 1000$

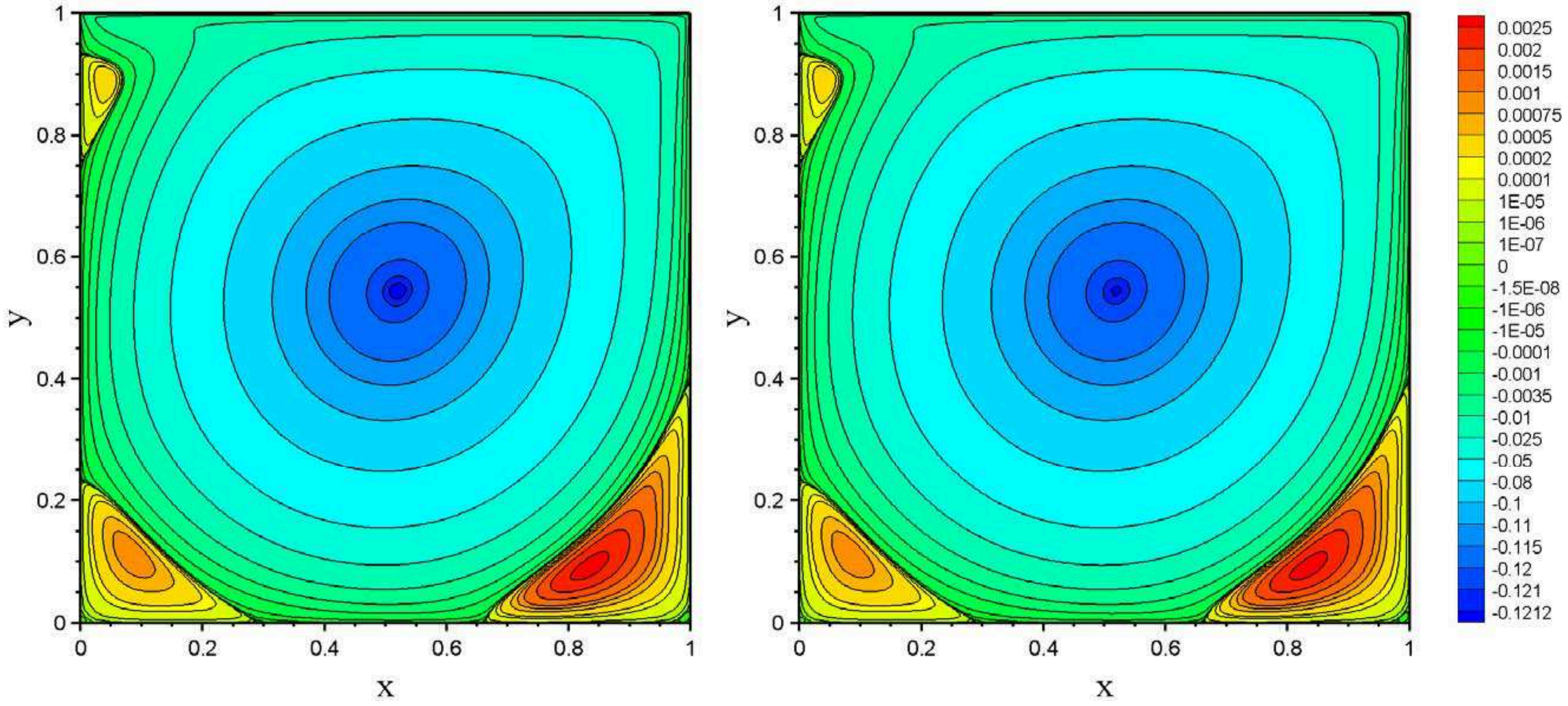

a) Compact solution

b) Wide solution

Figure 5) Fourth order streamfunction contours of driven cavity flow, $Re = 2500$

| | No. of points | $E_\psi$ | $m$ | $E_\omega$ | $m$ |
|---|---|---|---|---|---|
| compact formulation | 16×16 | 6.8849E-09 | | 1.3766E-08 | |
| | 32×32 | 4.5836E-10 | 3.91 | 9.1590E-10 | 3.91 |
| | 64×64 | 2.9545E-11 | 3.96 | 5.9001E-11 | 3.96 |
| | 128×128 | 1.9523E-12 | 3.92 | 3.8990E-12 | 3.92 |
| | 256×256 | 1.2396E-13 | 3.98 | 2.4502E-13 | 3.99 |
| wide formulation | 16×16 | 2.6410E-08 | | 5.0351E-08 | |
| | 32×32 | 1.7978E-09 | 3.88 | 3.5445E-09 | 3.83 |
| | 64×64 | 1.1710E-10 | 3.94 | 2.3164E-10 | 3.94 |
| | 128×128 | 7.7605E-12 | 3.92 | 1.5395E-11 | 3.91 |
| | 256×256 | 4.8750E-13 | 3.99 | 9.7813E-13 | 3.98 |

Table 1) Average absolute error and convergence rate of fourth order formulations

| *Re* | Ref. | $\psi$ | $\omega$ | $x$ | $y$ |
|---|---|---|---|---|---|
| 2500 | Present – Compact solution | -0.1212883 | -1.973684 | 0.5195 | 0.5430 |
| | Present – Wide solution | -0.1212327 | -1.972586 | 0.5195 | 0.5430 |
| | Present – Richardson extrap. sol. | -0.1214392 | -1.975666 | 0.5195 | 0.5430 |
| | Erturk and Gokcol [7] | -0.121472 | -1.976132 | 0.5200 | 0.5433 |
| | Barragy and Carey [2] | -0.1214621 | – | 0.5189 | 0.5434 |
| | Erturk, Corke and Gokcol [6] | -0.121470 | -1.976117 | – | – |
| 1000 | Present – Compact solution | -0.1188756 | -2.066955 | 0.5313 | 0.5664 |
| | Present – Wide solution | -0.1188513 | -2.066586 | 0.5313 | 0.5664 |
| | Present – Richardson extrap. sol. | -0.1189358 | -2.067704 | 0.5313 | 0.5664 |
| | Erturk and Gokcol [7] | -0.118938 | -2.067760 | 0.5300 | 0.5650 |
| | Botella and Peyret [3] | -0.1189366 | -2.067753 | 0.5308 | 0.5652 |
| | Schreiber and Keller [16] | -0.118821 | -2.0677 | – | – |
| | Erturk, Corke and Gokcol [6] | -0.118942 | -2.067213 | – | – |
| | Barragy and Carey [2] | -0.11893 | – | – | – |
| 100 | Present – Compact solution | -0.1035173 | -3.181031 | 0.6172 | 0.7383 |
| | Present – Wide solution | -0.1035161 | -3.181007 | 0.6172 | 0.7383 |
| | Present – Richardson extrap. sol. | -0.1035190 | -3.181046 | 0.6172 | 0.7383 |

Table 2) Properties at the center of the main circulating eddy in driven cavity flow

| $Re$ | $\alpha$ | Compact formulation | | Wide formulation | | $\frac{\text{CPU}_{\text{compact}}}{\text{CPU}_{\text{wide}}}$ | $\frac{\text{Iteration no}_{\text{compact}}}{\text{Iteration no}_{\text{wide}}}$ | $\left.\frac{\text{CPU}_{\text{compact}}}{\text{CPU}_{\text{wide}}}\right\|_{\max \Delta t}$ |
|---|---|---|---|---|---|---|---|---|
| | | $\text{CPU}_{\text{compact}}$ | $\text{Iteration no}_{\text{compact}}$ | $\text{CPU}_{\text{wide}}$ | $\text{Iteration no}_{\text{wide}}$ | | | |
| 2500 | 1.3 | 341.2 | 15937 | | | | | 1.33 |
| | 1.2 | 368.1 | 17211 | | | | | |
| | 1.1 | 398.9 | 18715 | 256.3 | 18482 | 1.56 | 1.01 | |
| | 1.0 | 439.0 | 20517 | 283.0 | 20269 | 1.55 | 1.01 | |
| | 0.9 | 486.7 | 22712 | 314.0 | 22464 | 1.55 | 1.01 | |
| | 0.8 | 546.6 | 25446 | 351.9 | 25200 | 1.55 | 1.01 | |
| | 0.7 | 625.1 | 28944 | 398.8 | 28706 | 1.57 | 1.01 | |
| | 0.6 | 719.4 | 33577 | 463.4 | 33358 | 1.55 | 1.01 | |
| | 0.5 | 858.0 | 40007 | 554.8 | 39937 | 1.55 | 1.00 | |
| 1000 | 1.3 | 326.7 | 16529 | | | | | 1.23 |
| | 1.2 | 350.6 | 17195 | | | | | |
| | 1.1 | 389.1 | 18704 | 266.2 | 19018 | 1.46 | 0.98 | |
| | 1.0 | 427.5 | 20507 | 291.9 | 20853 | 1.46 | 0.98 | |
| | 0.9 | 474.8 | 22702 | 322.1 | 23085 | 1.47 | 0.98 | |
| | 0.8 | 527.4 | 25430 | 360.2 | 25861 | 1.46 | 0.98 | |
| | 0.7 | 598.3 | 28915 | 410.3 | 29407 | 1.46 | 0.98 | |
| | 0.6 | 694.0 | 33523 | 478.3 | 34095 | 1.45 | 0.98 | |
| | 0.5 | 833.0 | 40048 | 568.1 | 40587 | 1.47 | 0.99 | |
| 100 | 1.3 | 242.4 | 11246 | | | | | 1.31 |
| | 1.2 | 255.7 | 12083 | | | | | |
| | 1.1 | 275.4 | 13172 | 185.5 | 13052 | 1.48 | 1.01 | |
| | 1.0 | 302.9 | 14478 | 202.8 | 14320 | 1.49 | 1.01 | |
| | 0.9 | 336.2 | 16071 | 224.8 | 15883 | 1.50 | 1.01 | |
| | 0.8 | 379.0 | 18059 | 251.7 | 17832 | 1.51 | 1.01 | |
| | 0.7 | 431.8 | 20610 | 287.5 | 20329 | 1.50 | 1.01 | |
| | 0.6 | 497.5 | 24003 | 336.1 | 23646 | 1.48 | 1.02 | |
| | 0.5 | 596.5 | 28739 | 401.8 | 28264 | 1.48 | 1.02 | |

Table 3) Numerical performances of compact and wide fourth order formulations with $128\times128$ grid mesh

| $Re$ | $\alpha$ | Compact formulation | | Wide formulation | | $\frac{CPU_{compact}}{CPU_{wide}}$ | $\frac{\text{Iteration no}_{compact}}{\text{Iteration no}_{wide}}$ | $\left.\frac{CPU_{compact}}{CPU_{wide}}\right\|_{\max \Delta t}$ |
|---|---|---|---|---|---|---|---|---|
| | | $CPU_{compact}$ | Iteration $no_{compact}$ | $CPU_{wide}$ | Iteration $no_{wide}$ | | | |
| 2500 | 1.3 | 6156.6 | 65039 | | | | | 1.36 |
| | 1.2 | 6700.5 | 69776 | | | | | |
| | 1.1 | 7253.1 | 76038 | 4525.2 | 72189 | 1.60 | 1.05 | |
| | 1.0 | 7961.0 | 83537 | 4975.3 | 79198 | 1.60 | 1.05 | |
| | 0.9 | 8738.5 | 92687 | 5500.3 | 87725 | 1.59 | 1.06 | |
| | 0.8 | 9770.7 | 104102 | 6076.6 | 98334 | 1.61 | 1.06 | |
| | 0.7 | 10998.9 | 118737 | 6834.0 | 111898 | 1.61 | 1.06 | |
| | 0.6 | 12838.1 | 138195 | 8053.0 | 129855 | 1.59 | 1.06 | |
| | 0.5 | 15608.1 | 165325 | 9784.8 | 154981 | 1.60 | 1.07 | |
| 1000 | 1.3 | 5894.5 | 62087 | | | | | 1.29 |
| | 1.2 | 6210.2 | 67136 | | | | | |
| | 1.1 | 5054.0 | 54731 | 3523.8 | 55593 | 1.43 | 0.98 | |
| | 1.0 | 5454.8 | 60109 | 3791.0 | 61454 | 1.44 | 0.98 | |
| | 0.9 | 6063.0 | 66665 | 4196.1 | 67659 | 1.44 | 0.99 | |
| | 0.8 | 6781.8 | 74835 | 4674.3 | 76031 | 1.45 | 0.98 | |
| | 0.7 | 7703.7 | 85302 | 5316.6 | 86771 | 1.45 | 0.98 | |
| | 0.6 | 8941.5 | 99195 | 6211.5 | 101055 | 1.44 | 0.98 | |
| | 0.5 | 10747.7 | 118538 | 7434.5 | 120988 | 1.45 | 0.98 | |
| 100 | 1.3 | 4534.2 | 46785 | | | | | 1.28 |
| | 1.2 | 4758.7 | 50239 | | | | | |
| | 1.1 | 5054.0 | 54731 | 3523.8 | 55593 | 1.43 | 0.98 | |
| | 1.0 | 5454.8 | 60109 | 3791.0 | 61454 | 1.44 | 0.98 | |
| | 0.9 | 6063.0 | 66665 | 4196.1 | 67659 | 1.44 | 0.99 | |
| | 0.8 | 6801.8 | 74835 | 4674.3 | 76031 | 1.46 | 0.98 | |
| | 0.7 | 7743.7 | 85302 | 5316.6 | 86771 | 1.46 | 0.98 | |
| | 0.6 | 8941.5 | 99195 | 6211.5 | 101055 | 1.44 | 0.98 | |
| | 0.5 | 10747.7 | 118538 | 7434.5 | 120988 | 1.45 | 0.98 | |

Table 4) Numerical performances of compact and wide fourth order formulations with $256\times256$ grid mesh